\RequirePackage{fix-cm}
\documentclass[twocolumn]{svjour3}          
\smartqed  
\usepackage{graphicx}
\usepackage{bm}
\usepackage{amsmath}
\usepackage{amssymb}
\usepackage{xcolor}
\usepackage{ragged2e}
\usepackage{mathptmx}      
\allowdisplaybreaks
\begin{document}

\title{Dynamics of Minimal Networks of Limit Cycle Oscillators}


\author{Andrea Elizabeth Biju       \and
        Sneha Srikanth              \and
        Krishna Manoj               \and
        Samadhan A. Pawar           \and
        R. I. Sujith
}


\institute{ A. E. Biju$^{1, 2}$ \and 
            S. Srikanth$^3$ \and K. Manoj$^4$ \and S. A. Pawar$^5$ \and R. I. Sujith$^{1,2,\star}$ \at
            $^1$Center of Excellence for Studying Critical Transitions in Complex Systems, Indian Institute of Technology Madras, Chennai 600036, India \\
            $^2$Department of Aerospace Engineering, Indian Institute of Technology Madras, Chennai 600036, India \\
            $^3$Department of Mechanical Engineering, Indian Institute of Technology Madras, Chennai 600036, India \\
            $^4$Department of Mechanical Engineering, Massachusetts Institute of Technology, Cambridge, MA 02139, USA\\
            $^5$Institute for Aerospace Studies, University of Toronto, Toronto, Ontario M3H 5T6, Canada\\
           \email{$^{\star}$sujith@iitm.ac.in}
}

\date{Received: date / Accepted: date}

\maketitle

\begin{abstract}
The framework of mutually coupled oscillators on a network has served as a convenient tool for investigating the impact of various parameters on the dynamics of real-world systems. Compared to large networks of oscillators, minimal networks are more susceptible to changes in coupling parameters, the number of oscillators, and network topologies. In this study, we systematically explore the influence of these parameters on the dynamics of a minimal network comprising Stuart-Landau oscillators coupled with a distance-dependent time delay. We examine three network topologies: ring, chain, and star. Specifically, for ring networks, we study the effects of increasing nonlocality from local to global coupling on the overall dynamics of the system. Our findings reveal the existence of various synchronized states, including splay and cluster states, a partially synchronized state such as chimeric quasiperiodicity, and an oscillation quenching state such as amplitude death in these networks. Moreover, through an analysis of long-lived transients, we discover novel amplitude-modulated states within ring networks. Interestingly, we observe that increasing nonlocality diminishes the influence of the number of oscillators on the overall behavior in these networks. Furthermore, we note that chain networks, unlike ring networks, do not exhibit perfect synchrony among the coupled oscillators. In contrast, star networks demonstrate greater stability and are unaffected by the number of oscillators within the network. The insights from this study deepen our understanding of the dynamics of minimal networks and have implications for various fields, ranging from biology to engineering.
\end{abstract}

\section{Introduction}
\label{sec1}
\sloppy
Understanding and controlling the dynamical behavior of physical systems, both natural and human-made, have long intrigued the scientific community and continue to be a topic of immense interest \cite{lehnert_controlling_2016,pikovsky_synchronization:_2007,strogatz_sync:_2003}.  Ranging from the interaction of large groups of animals \cite{sumpter_principles_2006} to social networks \cite{ubaldi_emergence_2021}, various collective oscillatory dynamics emerge due to complex nonlinear interactions between the individual units \cite{raducha_emergence_2020}. These dynamical states include synchronization \cite{pikovsky_synchronization:_2007}, quenching of oscillations \cite{zou_quenching_2021}, and partially synchronized states such as chimera \cite{haugland_between_2021,zakharova_chimera_2020} and weak chimera \cite{ashwin_weak_2015}. Such dynamical behavior stems from the interplay of several factors, including the coupling scheme, topology, strength of the interaction, and the number of interacting subunits in the system \cite{dixit_emergent_2021}.

Interestingly, the emergence of collective dynamical behavior can be observed in systems with varying levels of complexity -- ranging from large systems with thousands of interacting components, such as brain networks \cite{ma_review_2015} and ant colonies \cite{chialvo_how_1995}, to minimal systems with only a few (i.e., 3 to 20) oscillators, such as players co-ordinating in a team sport \cite{yokoyama_three_2011}, limb co-ordination while walking \cite{getchell_developmental_2007}, coupled lasers \cite{soriano_complex_2013}, and gas turbine combustors \cite{guan_synchronization_2022,moon_modal_2023}. In large systems, the addition or removal of a few oscillators, or alteration of their connectivity rarely affects the overall dynamics of the system \cite{arenas_synchronization_2008}. However, the same cannot be said for systems with a minimal number of oscillators. 

For instance, Manoj \textit{et al.} \cite{manoj_experimental_2018,manoj_synchronization_2019} showed that increasing the number of oscillators from two to four resulted in an increase of the variety of the dynamical states using experiments on candle oscillators. They also observed that rings of coupled candle oscillators with even number of oscillators show anti-phase clustering, while rings with an odd number of oscillators did not. This observation was later shown by Joseph \textit{et al.} \cite{vathakkattil_joseph_limits_2020} to be exclusive to minimal networks with less than 20 oscillators. Change of network topology can also significantly influence the dynamics of minimal networks. In this context, Manoj \textit{et al.} observed that in candle oscillators, phase synchronization is only observed in closed-loop topologies \cite{manoj_experimental_2021}. Furthermore, Zou and Zhan \cite{zou_splay_2009} showed that the stability of various synchronized states observed in ring networks change as the coupling scheme is varied from local to global. Moreover, to reflect the finite time needed for information to propagate between oscillators in real-life systems. studies have included time-delay in the coupling \cite{lakshmanan_dynamics_2011}. Introducing these time-delays can also lead to changes in the stability of the observed dynamical states \cite{ramana_reddy_time_1998}. Through these representative examples, we understand that the network topology, coupling scheme, number of oscillators and time-delays play a crucial role in determining the dynamical behavior of a minimal network of oscillators.

Although the aforementioned studies and several others have separately considered the effects of time-delay \cite{ramana_reddy_time_1998}, coupling schemes \cite{zou_splay_2009}, the number of oscillators \cite{ramana_reddy_time_1998,dodla_phase-locked_2004,mehta_amplitude_2006} and topology \cite{ashwin_weak_2015} on the dynamics of small networks of coupled limit cycle oscillators, there have been very few attempts at performing a systematic analysis of the effects of all of these properties within the same system. Furthermore, the vast majority, if not all, of these studies ignore long transient dynamics. However, the information given by the transients can be important in some cases, such as the population dynamics of species in an ecosystem \cite{defriez_climate_2016} and combustors undergoing thermoacoustic instability \cite{roy_flame_2021}. In fact, certain transients possess interesting properties and can even last for uncharacteristically long times, with their lifetime depending on the number of oscillators \cite{wolfrum_chimera_2011} and coupling parameters.

In the current study, we investigate the dynamics of a minimal network of Stuart-Landau oscillators to understand the effect of varying the number of oscillators, time-delay, coupling scheme, and network topology in the same system. We comprehensively identify all possible dynamical states and characterize the long transient behavior manifested in the coupled oscillator system. We observe that an increase in nonlocality can significantly alter the variety of dynamical states in a ring of oscillators. We also demonstrate that the dynamics of a minimal network depend critically on the topology of the connections by studying the variations across ring, star, and chain topologies.

The rest of the paper is organized as follows: Section~\ref{Model} presents the mathematical details of the model and the coupling structures investigated in the study. Section~\ref{results} presents the results, where we first enlist and describe the various types of dynamical states observed in the study (Sec.~\ref{states}). Then, we examine how varying the different system and coupling parameters affect the occurrence of these states in ring networks (Sec.~\ref{ring}). Next, we compare these results with the dynamical behavior of chain and star networks (Sec.~\ref{chain_star}). Finally, we conclude our study and highlight its possible applications in Section~\ref{conclusion}.

\section{\label{Model}Model}
\sloppy
To analyze the dynamical behavior of minimal networks under various topologies and coupling schemes, we construct a model using time-delay coupled Stuart-Landau oscillators. The Stuart-Landau oscillator is a well-known mathematical model that has been extensively utilized to capture the phase and amplitude dynamics of limit cycle oscillations in different fields of science and engineering \cite{frasca_synchronization_2018,garcia-morales_complex_2012,lakshmanan_dynamics_2011,atay_amplitude_2009,zakharova_chimera_2020}. Similarly, coupled Stuart-Landau oscillators have been used to model several systems such as coupled lasers \cite{soriano_complex_2013}, turbulent flows \cite{callaham_empirical_2022,thompson_stuartlandau_2004,haugland_between_2021}, brain networks \cite{moon_general_2015}, candle flame oscillators \cite{manoj_experimental_2018}, chemical oscillators \cite{zou_revival_2017}, and thermoacoustic oscillators \cite{premraj_effect_2021}. The equation of a Stuart-Landau oscillator is given by:
\begin{equation}
    \Dot{z}(t) = (a + i\omega_n - |z(t)|^2) z(t),
    \label{eq1}
\end{equation}
where $z=x+iy$ is a complex variable, $a$ is the Hopf bifurcation parameter, and $\omega_n$ is the natural frequency of the oscillator. The above equation has a fixed point at the origin that is stable for $a<0$ and unstable for $a>0$, in which limit cycle oscillations are observed via supercritical Hopf bifurcation. For simplicity, we consider that all oscillators are identical, i.e., $a$ and $\omega_n$ are the same for all oscillators.

The simplest minimal network is the case of two time-delay coupled Stuart-Landau oscillators, the behavior of which has been well-investigated in literature \cite{aronson_amplitude_1990,ramana_reddy_time_1998,atay_amplitude_2009}. However, when the number of oscillators $N$ is further increased, there are multiple topologies and coupling schemes in which the oscillators can be coupled \cite{lakshmanan_dynamics_2011}. 
Several studies \cite{dodla_phase-locked_2004,mehta_amplitude_2006,vathakkattil_joseph_limits_2020} indicate that the dynamics of the local and nonlocal ring networks become independent of the number of oscillators for $N>20$. Hence, to systematically analyze the sensitivity of minimal networks to changes in coupling parameters, we consider networks with $N<20$, specifically $N=6$ to $9$. The oscillators in a minimal network can be coupled in several topologies. In this study, we first consider the simplest case of a ring network. We explore the effect of varying the coupling scheme from local to global via nonlocal coupling. We then compare the case of ring networks to open network topologies such as star and chain networks.

\subsection{Ring topology}
The general equation for a ring of identical delay coupled Stuart-Landau oscillators is given by:
\begin{align}
\begin{split}
    \Dot{z}_j(t) &= (a + i\omega - |z_j(t)|^2) z_j(t) \\ 
    &\quad+ \frac{K}{p} \sum_{l = 1}^{p} [z_{j-l}(t - \tau_l) + z_{j+l}(t - \tau_l) - 2z_j(t)] \label{eq2}
\end{split}
\end{align}
Periodic boundary conditions are employed in this network, specified as $z_{N+j} = z_j$ and $z_{-j} = z_{N-j}$. Here, $N$ represents the number of oscillators in the network, $K$ is the coupling strength, and $p$ is the coupling range, i.e., the number of oscillators to which the $j^\textrm{th}$ oscillator is connected on either side, with $j=1, ..., N$. We assume the time-delay $\tau_l$ in the coupling of oscillators as:
\begin{equation}
    \tau_l = 2r \sin\bigg(\frac{l\pi}{N}\bigg),
    \label{eq3}
\end{equation}
where $r$ is the radius of the ring. Equation (\ref{eq3}) implies that the time-delay is proportional to the chord length between the connected oscillators. This expression for time-delay becomes important in systems such as thermoacoustic systems \cite{dange_oscillation_2019,moon_cross-talk-interaction-induced_2020}, coupled candle flames \cite{manoj_experimental_2018}, and metronomes \cite{dudkowski_experimental_2016}, where the interaction distance is utilized to realize the coupling between two oscillators. Here, the shortest distance between the oscillators determines the time-delay between them.

In the case of local coupling, each oscillator is connected to its nearest neighbors, i.e., $p=1$. Equation (\ref{eq2}) then becomes:
\begin{align}
\begin{split}
    \Dot{z}_j(t) &= (a + i\omega - |z_j(t)|^2) z_j(t) \\
    &\quad+ K[z_{j+1}(t - \tau) + z_{j-1}(t - \tau) - 2z_j(t)],
    \label{eq4}
\end{split}
\end{align}
where $\tau = 2r \sin(\pi/N)$, and is identical for all links between the oscillators.

Nonlocal coupling corresponds to the case when each oscillator is connected to more than one oscillator beyond its nearest neighbors. For simplicity, we assume that every oscillator is connected to the same number of neighbors on each side. For such a coupling scheme, the equation governing the $j^\textrm{th}$ oscillator dynamics is given by equation (\ref{eq2}), where $p=1, 2, ..., (N-1)/2$ for all combinations of $N$ and $p$. However, when $N$ is an even number and $p=(N-1)/2$, a different equation is used, given by:
\begin{align}
\begin{split}
    \Dot{z}_j(t) &= (a + i\omega - |z_j(t)|^2) z_j(t) \\
    &\quad+ \frac{K}{p} \sum_{l = 1}^{\lfloor{p}\rfloor} [z_{j-l}(t - \tau_l) + z_{j+l}(t - \tau_l) - 2z_j(t)]\\ 
    &\quad+ \frac{K}{p} [z_{N/2}(t - \tau_{N/2}) - z_j(t)],
\label{eq5}
\end{split}
\end{align}
where $\lfloor{p}\rfloor$ corresponds to the floor of $p$.

Each oscillator in a globally coupled ring is connected to all the other oscillators, but with the same coupling strength and delay. The equation for the $j^\textrm{th}$ oscillator in a globally coupled ring is given by:
\begin{align}
\begin{split}
    \Dot{z}_j(t) &= (a + i\omega - |z_j(t)|^2) z_j(t)\\ 
    &\quad+ \frac{2K}{N-1} \sum_{l = 1, \: l\neq j}^{N} [z_{l}(t - \tau) - z_j(t)],
    \label{eq6}
\end{split}
\end{align}
where $\tau = 2r \sin(\pi/N)$, identical for all links in the network.

\subsection{Chain and star topologies}
In order to further analyze the effect of topology on the dynamical behavior of the network of oscillators, we compare the dynamics of the ring of oscillators to two of the simplest cases of open network topologies, i.e., chain and star networks. 

A chain network is simply a locally coupled ring network as represented by equation~(\ref{eq4}), except that the ends are not connected, i.e., the boundary conditions are $z_{N+1} = 0$ and $z_{0} = 0$. In a star network, the center oscillator is assigned as the `hub', and the other oscillators are considered the `leaves'. The equation for the hub oscillator (say, oscillator 1) in a start network of coupled Stuart-Landau oscillators is:
\begin{equation}
\begin{split}
    \Dot{z}_1(t) &= (a_1 + i\omega_1 - |z_1(t)|^2) z_1(t)\\ 
    &\quad+ K\sum_{j=2}^N[z_{j}(t - \tau) - z_1(t)]
    \label{eq7}
\end{split}
\end{equation}
while the equation for all other oscillators on the leaves of the network is:
\begin{equation}
\begin{split}
    \Dot{z}_j(t) &= (a_j + i\omega_j - |z_j(t)|^2) z_j(t)\\ 
    &\quad+ K[z_{1}(t - \tau) - z_j(t)]
    \label{eq8}
\end{split}
\end{equation}
for all $j=2,3, ..., N$. We note that oscillators on different leaves in the star network do not directly interact with each other; they interact via the hub node.

We numerically integrate the equations using the fourth-order Runge-Kutta method in MATLAB (2022b), with a time step of 0.01. We fix the Hopf bifurcation parameter $a$ as 1 and the natural frequency $\omega_n$ as 10 (Eq.~\ref{eq1}) for all identical oscillators. We sweep through the values of $r$ and $\tau$ in steps of 0.01 in our simulations. We notice complex dynamical behavior in ring networks for lower coupling strength ($K\leq1$) as compared to higher coupling strength ($K>1$). For this reason, we vary the coupling strength in finer steps of 0.1 for $K\leq1$, and in coarse steps of 1 for $K>1$. The initial conditions for oscillators are equally spaced between $[-1, 1]$ in all cases. The first 5000 time steps are removed to eliminate the usual transient behavior. However, uncharacteristically long transients are examined in detail. 

\section{\label{results}Results and discussion}

In this section, we discuss the behavior of a minimal network of time-delay coupled Stuart-Landau oscillators in different topologies. First, we classify the variety of global dynamical states, both stable and long transients, observed at different coupling conditions of networks. Subsequently, we analyze and compare the similarities and differences in the collective behaviors exhibited by ring, chain, and star networks.

\subsection{\label{states}Dynamical states observed in ring, chain, and star networks}

A network of coupled Stuart-Landau oscillators is capable of exhibiting a wide variety of dynamical states, including various states of synchronization \cite{boccaletti_synchronization:_2018,osipov_synchronization_2007}, chimera \cite{zakharova_chimera_2020}, and oscillation quenching \cite{zou_quenching_2021}. The characterization of these states depends on the properties of frequency, phase, and amplitude locking of coupled oscillators. Here, we use the frequency, phase, and amplitude tolerances to distinguish various states as $\pm 0.01$, $\pm 1^{\circ}$, and $\pm$ 0.01, respectively.

\begin{figure*}
    \centering
    \includegraphics[width = 0.95\textwidth]{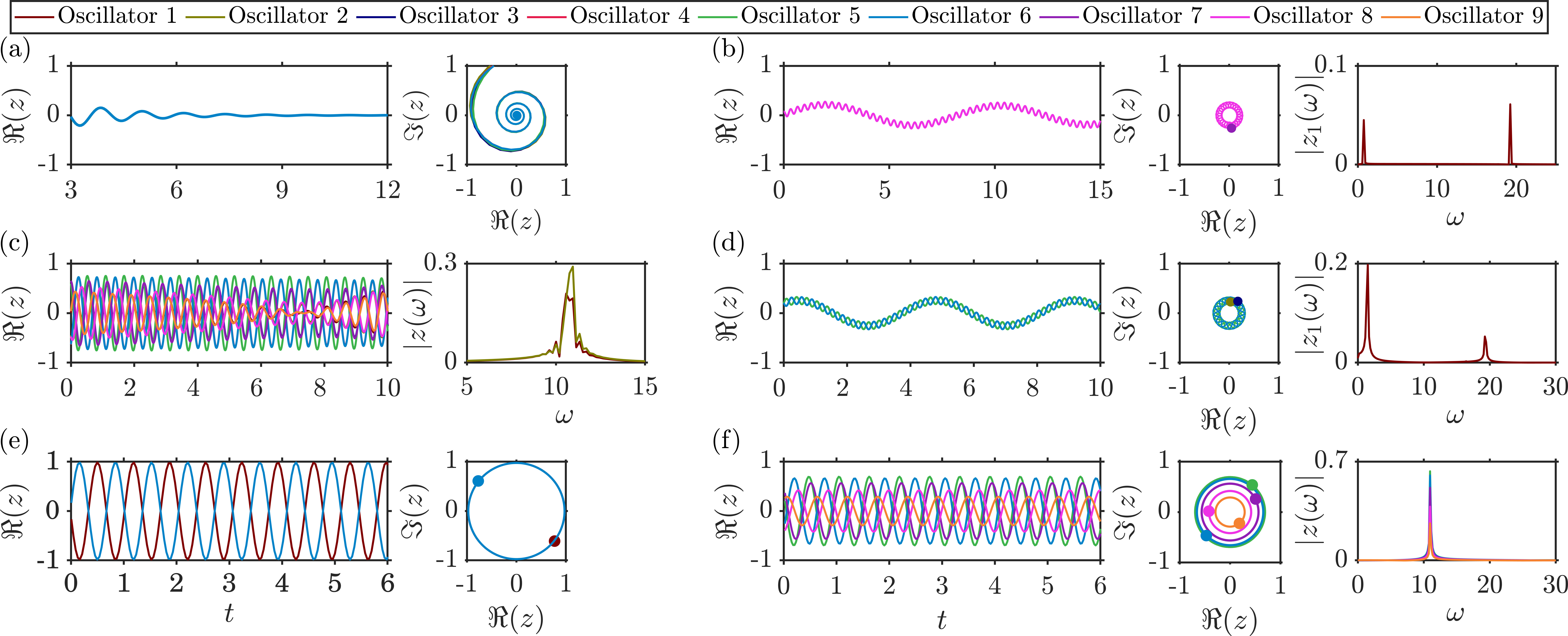}
    \caption{Various dynamical states observed in the network of coupled Stuart-Landau oscillators. (a) Amplitude death (ring network, $N=6$, local coupling, $K=3$, $r=0.15$; time series and phase space representation shown). (b) Amplitude-modulated in-phase (AMIP) state (ring network, $N=8$, global coupling, $K=19.5$, $r=0.41$; time series, phase space representation and amplitude spectrum shown). (c) Chimeric quasiperiodicity (CQP) (chain network, $N=9$, $K = 0.6$, $\tau = 0.20$; time series and amplitude spectrum of two oscillators shown). (d) Amplitude-modulated 2-cluster (AM2CL) state (ring network, $N=6$, local coupling, $K=25$, $r=0.15$; time series, phase space representation and amplitude spectrum shown). (e) Anti-phase remote synchronized (APRS) state (star network, $N=6$, $K=10$, $\tau = 0.35$; time series and phase space representation shown). (f) 5 amplitude cluster (5ACL) state (chain network, $N=9$, $K=0.7$, $\tau=0.20$; time series, phase space representation and amplitude spectrum of all oscillators shown).}
    \label{fig1}
\end{figure*}

\begin{figure*}
    \centering
    \includegraphics[width = 0.95\textwidth]{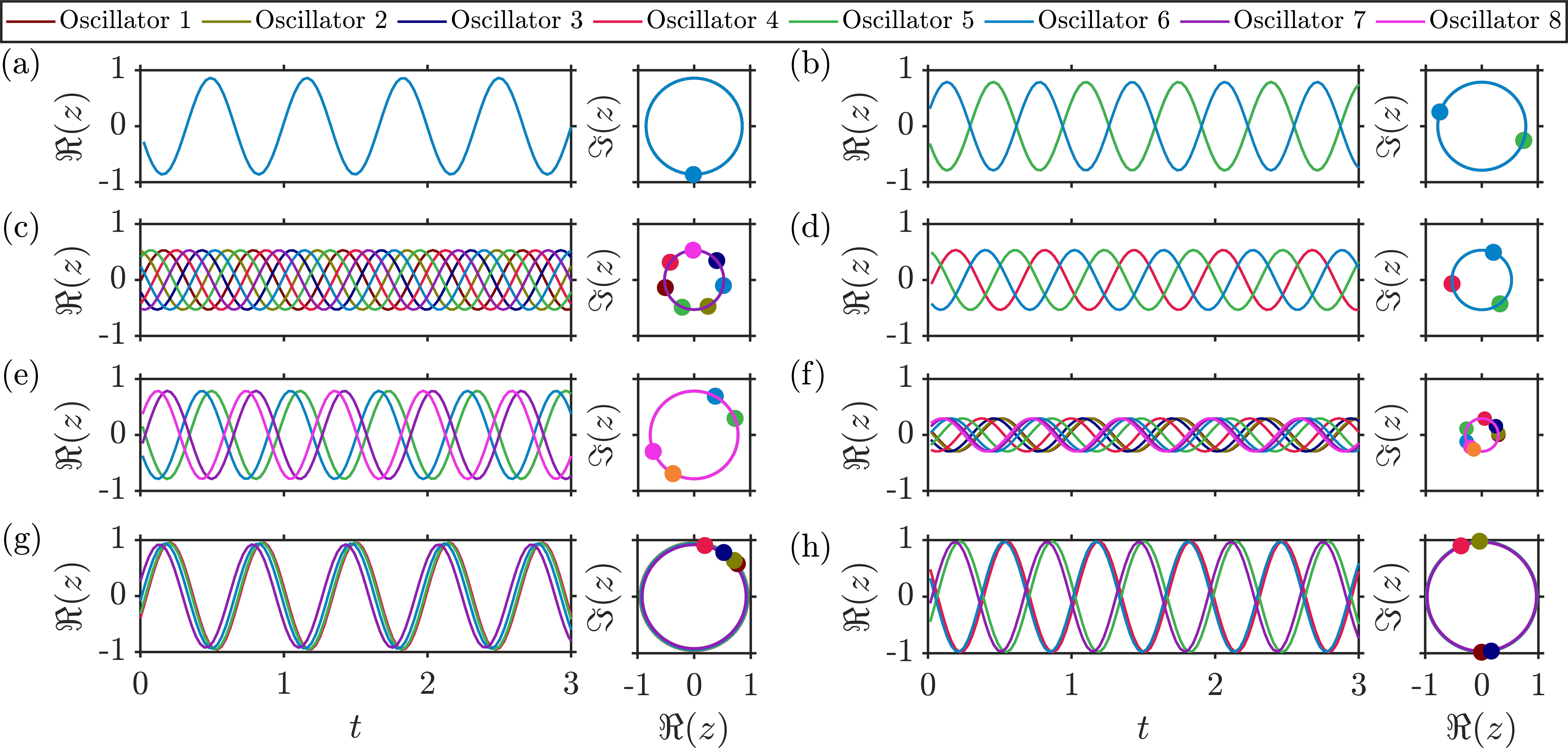}
    \caption{Time series and phase-space representation of different states of synchronization in a minimal network. (a) In-phase synchronization (ring network, $N=6$, local coupling, $K =25$, $r=0.05$), (b) 2-cluster state (ring network, $N=6$, local coupling, $K=25$, $r = 0.30$), (c) splay state of type 1 (ring network, $N=7$, nonlocal coupling, $K=0.5$, $r = 0.15$), (d) 3-cluster state (ring network, $N=6$, nonlocal coupling, $K=0.5$, $r=0.25$), (e) generalized 4-cluster state (ring network, $N=8$, nonlocal coupling, $K=0.3$, $r=0.17$), (f) generalized splay state (ring network, $N=8$, global coupling, $K=0.5$, $r=0.30$), (g) near in-phase state (chain network, $N=7$, $K=0.4$, $\tau = 0.05$), (h) near 2-cluster state (chain network, $N=7$, $K=0.4$, $\tau=0.35$).}
    \label{fig2}
\end{figure*}

In Figs. \ref{fig1} and \ref{fig2}, we show the various dynamical states observed across different networks discussed in Sec. \ref{ring} and \ref{chain_star}. The state of amplitude death (AD) is observed in all network topologies for intermediate values of time-delay and coupling strength between oscillators (Fig.~\ref{fig1}a). During this state, the amplitudes of all oscillators in the coupled system decay down to zero.  

In Fig. \ref{fig1}(b), we present different properties of the state of amplitude-modulated in-phase (AMIP) oscillations. This state is primarily observed as a transient at high values of coupling strength in locally coupled ring networks with an odd number of oscillators (Fig.~\ref{fig4}a,c), and globally coupled ring networks for any number of oscillators (Fig.~\ref{fig8}). During the occurence of this state, we notice two dominant frequency peaks that are far apart in the amplitude spectrum (\ref{fig1}b). The fast oscillation waveform of all the oscillators is modulated at a smaller frequency due to the nonlinear interaction between slow and fast timescales of the oscillators. We also witness a novel state of amplitude-modulated 2-cluster (AM2CL) in locally coupled ring networks with an even number of oscillators (Fig.~\ref{fig1}d). During this state, we notice a coexistence of in-phase and 2-cluster states in the global dynamics of the system, where in-phase synchrony is observed in the long time scales and anti-phase synchrony is observed in the short time scales.

Furthermore, we found that as the two frequencies observed in AMIP and AM2CL states are incommensurate, the corresponding oscillations exhibit the property of quasiperiodicity despite all oscillators being in synchrony. Hence, we refer to this state as synchronied quasiperiodic (SQP) oscillations. In constant, we also observe the state of transient chimeric quasiperiodicity (CQP) at lower coupling strengths, in both ring and chain networks (Fig.~\ref{fig1}c). During the CQP state, the amplitude spectrum of each oscillator is broadband, with multiple small peaks that are not necessarily the same for each oscillator. We also note that the CQP state is similar to the state of weak chimera observed in minimal networks \cite{ashwin_weak_2015}, where there are groups of frequency synchronized (i.e., having the same frequencies) and desynchronized (i.e., having the different frequencies) oscillators that coexist in the system. The occurrence of such states (i.e., SQP and CQP) indicates that coupling between oscillators in a network can lead to the emergence of multiple new frequencies, and hence can result in quasiperiodic behavior of individual oscillators. Note that, the quasiperiodic chimeras described in \cite{lee_chaotic_2023,pikovsky_partially_2008} are different from the state we describe here; the order parameter of the quasiperiodic chimera state exhibits in these studies a quasiperiodic breathing nature, which is not observed in the current study.

Moreover, we observe the state of anti-phase remote synchronization \cite{chen_stability_2022}, especially in star networks (Fig.~\ref{fig1}e). During this state, oscillators on the leaf nodes of the network are in-phase synchronized with each other while being in anti-phase synchronization with the oscillator at the hub node. An interesting behavior of amplitude cluster (ACL), similar to that observed in \cite{haugland_coexistence_2023}, is also witnessed in chain networks. Figure \ref{fig1}(f) highlights the properties of the state of five amplitude clusters observed in the system, where the dynamics of synchronized oscillators (i.e., having the same dominant frequency) are grouped into different amplitudes. 

In addition to the aforementioned complex dynamical behaviors in Fig. \ref{fig1}, we also notice a variety of splay states mostly in ring networks of oscillators (see Fig. \ref{fig2}). During a splay state, although all oscillators in a network are synchronized, the mean phase difference between them is not constant and is equally distributed on a ring, as given below:
\begin{equation}
    \theta_j = \omega t + (j-1)2\pi\frac{m}{N}
\end{equation}
where $j=1,2,..., N$, $\omega$ is the frequency of synchronized oscillations, $t$ is time, and $m$ is the wave number (or mode) of the splay state. Depending on the value of $m$, many varieties of synchronized oscillations in a splay state can be observed \cite{choe_controlling_2010}. 

\begin{figure}
    \centering
    \includegraphics[width = \columnwidth]{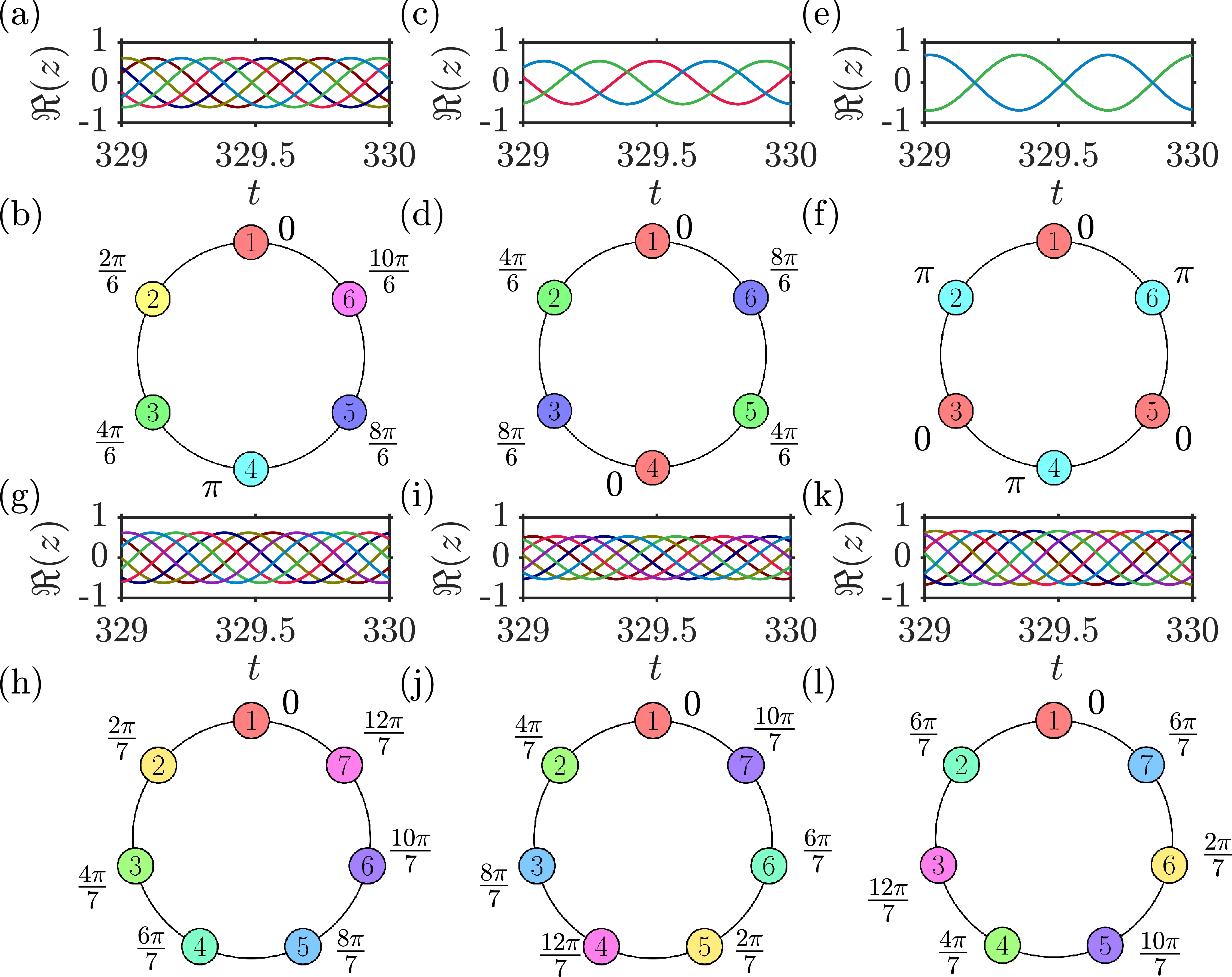}
    \caption{Time series and schematic representation of phases of nonlocally coupled oscillators on a ring network for splay states observed when (a-f) $N=6$ and (g-l) $N=7$ for $p=3$. (a-b) Splay state of type 1 ($m=1$, $r=0.15$), (c-d) 3-cluster state ($m=2$, $r=0.25$), and (e-f) 2-cluster state ($m=3$, $r=0.45$), (g-h) splay state of type 1 (mode $m=1$, $r=0.15$), (i-j) splay state of type 2 ($m=2$, $r=0.25$), (k-l) splay state of type 3 ($m=3$, $r=0.45$). For all cases, $K=0.5$.}
    \label{fig3}
\end{figure}

For $m=0$, we notice the state of in-phase synchronization between all $N$ oscillators, where they are locked at the zero phase difference (Fig~\ref{fig2}a). When $m=2$, we observe two groups of oscillators locked in anti-phase synchronization (i.e., the difference of $\pi$ rad) with each other. This state can also be referred to as 2 clusters (Fig~\ref{fig2}b). Similarly, we notice the occurrence of three groups (or 3 clusters) separated by $2 \pi/3$ rad of phase difference for $m=3$ (Fig~\ref{fig2}d). Thus, clustering is a subset of the splay state observed for different values of $m$ in the ring network \cite{choe_controlling_2010}. In Figs. ~\ref{fig3}(a-f) and ~\ref{fig3}(g-l), we show different types of splay states ($m>1$) observed in the case of $N=7$ and $N=6$ of ring networks, respectively. Depending on the mode ($m$) of the splay state, the mean phases of all oscillators on a ring are equally distributed between 0 to $2\pi$ rad (refer to Fig. ~\ref{fig2}c and Figs. ~\ref{fig3}). Only $m$ up to $\lfloor N/2 \rfloor$ gives unique splay states.

Furthermore, we observe other synchronized states such as the generalized splay and generalized cluster states in the ring network. A generalized splay state (Fig.~\ref{fig2}f), unlike the splay state discussed before, need not have the equal distribution of the phase difference between the oscillators (Fig. \ref{fig2}f); however, the center of mass, or effectively, the order parameter of the oscillators must be zero, similar to that of the splay state \cite{berner_generalized_2021,zou_splay_2009}. On the other hand, in the case of a generalized cluster state, the oscillators group into clusters, where the phase difference among clusters is not equal; however, the order parameter of all oscillators is observed to be equal to zero. The generalized 4-cluster state observed in the ring of eight oscillators is shown in Fig.~\ref{fig2}(e). A similar state has also been reported in experiments on eight locally coupled nano-electromechanical oscillators in \cite{matheny_exotic_2019}.

In a chain topology, we do not observe perfect phase locking into clusters; instead, we observe near in-phase synchrony (Fig. \ref{fig2}g) and near 2-cluster (Fig. \ref{fig2}h) states. We note that, in both these cases, oscillators (1, 7), (2, 6), (3, 5) and 4 are not exactly in-phase synchronized. However, these clusters are slightly phase shifted from each other (by less than $90^{\circ}$), and hence are not exactly in-phase (Fig. \ref{fig2}g) or anti-phase (Fig. \ref{fig2}h) synchronized to each other.

\subsection{\label{ring}Dynamical behavior of ring networks}

In this section, we characterize the dynamical states observed in a ring network of coupled Stuart-Landau oscillators by varying the nonlocality parameter $p$ in equation (\ref{eq2}). We also study the effect of variation of other parameters, such as the number of oscillators, coupling strength, and time-delay on the dynamics of such networks. We vary the time-delay between the oscillators in the network by changing the radius ($r$) of the ring as mentioned in equation (\ref{eq3}).

\subsubsection{Local coupling ($p=1$)}

\begin{figure*}
    \centering
    \includegraphics[width = 0.8\textwidth]{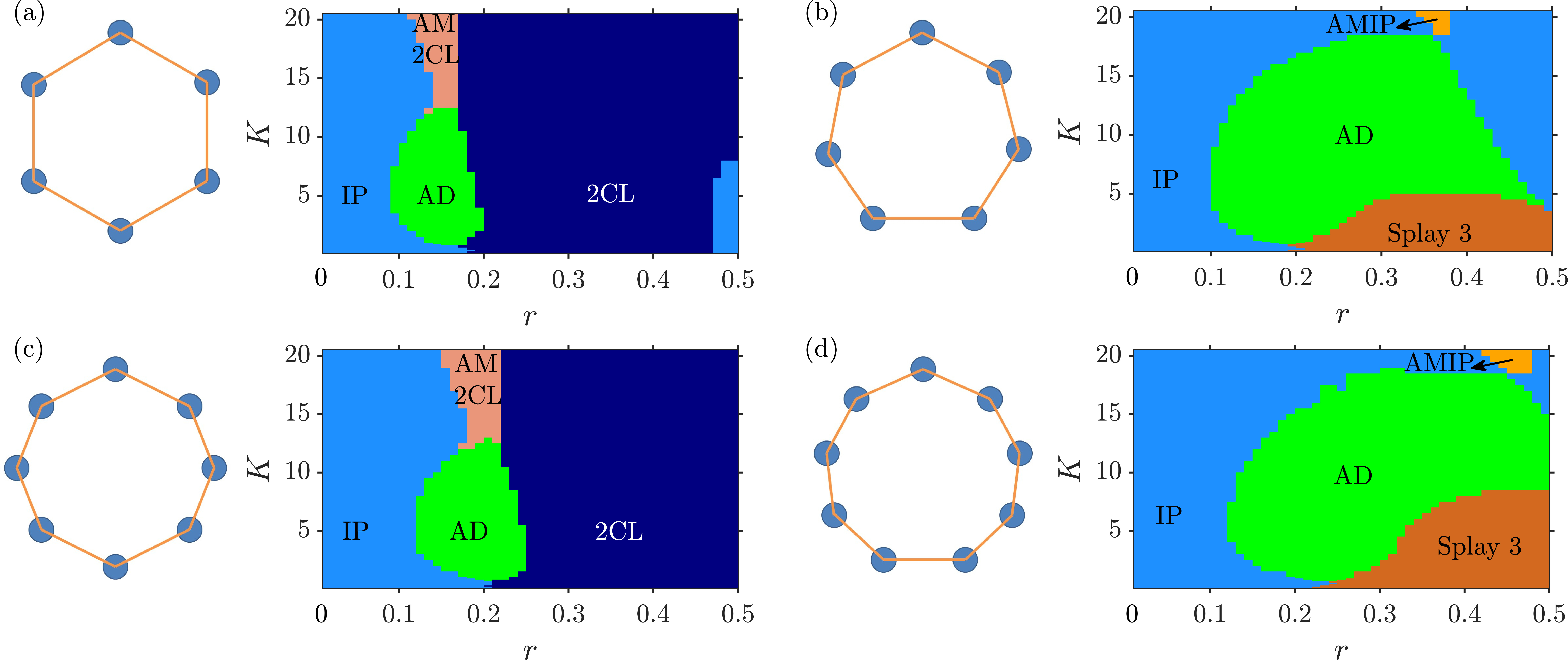}
    \caption{(a)-(d) Pictorial representation of the networks and the corresponding $K$-$r$ plots showing the dynamics of a locally coupled ring of oscillators for $N=$ 6, 7, 8, and $9$, respectively. Here, abbreviations IP, AD, 2CL, AM2CL, AMIP, and Splay 3 stand for in-phase oscillations, amplitude death, the 2-cluster state, the amplitude-modulated 2-cluster state, the amplitude-modulated in-phase state, and the splay state of type 3, respectively.}
    \label{fig4}
\end{figure*}

The overall dynamical behavior of rings of locally coupled oscillators is depicted using $K$-$r$ plots in Fig.~\ref{fig4}. For the case of an even number of oscillators in the ring, i.e., $N=6$ and $N=8$ (Figs.~\ref{fig4}a and \ref{fig4}c), we observe the prominent dynamical states as in-phase synchronization (IP), 2-cluster (2CL), amplitude death (AD), and amplitude-modulated 2-cluster (AM2CL). At lower coupling strengths ($K<1$), as the time-delay ($r$) increases, we notice a direct transition from an IP to a 2-CL state. At intermediate coupling strengths, the transition from IP to 2-CL state occurs via a region of AD. In contrast, at higher coupling strengths (around $K>12$), the transition occurs via a region of amplitude-modulated 2-cluster state (AM2CL). 

The $K$-$r$ plots for the case of odd numbers of oscillators (Figs. \ref{fig4}b and \ref{fig4}d) are remarkably different from that of the case with an even number of oscillators. At low values of coupling strengths ($K<1$), as the time-delay is increased, a direct transition between the IP state and splay state of the highest mode, which is the splay state of type 3 in the case of $N=7$ (Fig.~\ref{fig3}k, l) and $N=9$ oscillators, is observed. At intermediate coupling strengths, the transition between these states occurs via the region of AD. We observe only regions of IP and AD at sufficiently higher coupling strengths. Moreover, a short-lived state of amplitude-modulated in-phase (AMIP) oscillations, as well as stable in-phase oscillations are observed at very high coupling strengths. Furthermore, we note that the amplitude death region is also much larger for networks with an odd number of oscillators than in the case of an even number of oscillators on the ring (compare AD regions in Figs. \ref{fig4}a,c and \ref{fig4}b,d). Thus, we can quench oscillations (if undesired) easily over a wider range of coupling parameters if the number of oscillators in a ring is odd. Upon increasing the number of oscillators in the network ($N$), the amplitude death region expands and shifts toward a higher value of time-delay.

\subsubsection{Nonlocal coupling ($p=2$)}
\begin{figure*}
    \centering
    \includegraphics[width = 0.8\textwidth]{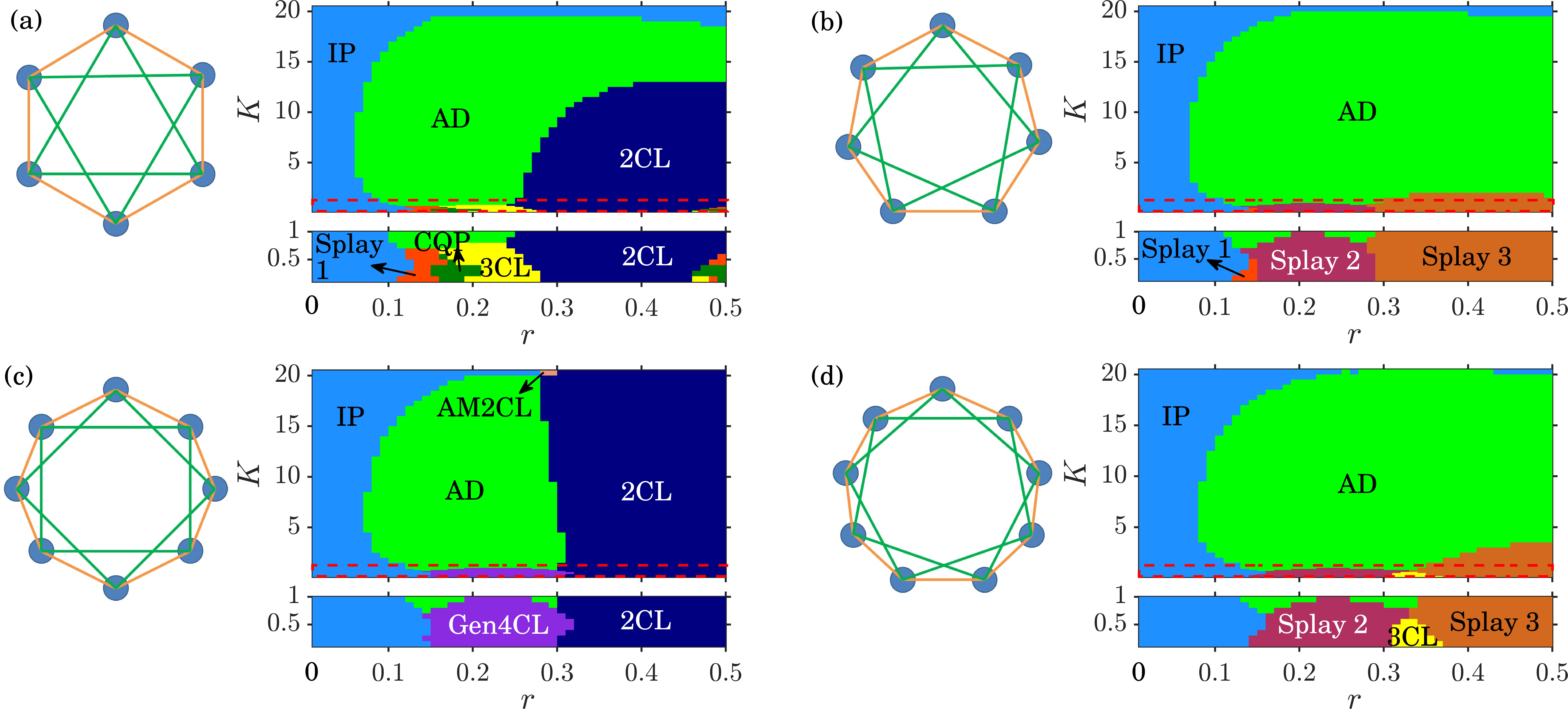}
    \caption{(a)-(d) Pictorial representation of the networks and the corresponding $K$-$r$ plots showing the dynamics of a nonlocally coupled ring of oscillators ($p=2$ case) for $N=6$, $7$, $8$, $9$, respectively. Zoomed inset of the region highlighted by the red dashed box is shown below ($K<1$ region). Here, the abbreviations 3CL, Splay 1, Splay 2, Gen4CL and CQP stand for 3-cluster state, splay state of type 1, splay state of type 2, generalized 4-cluster state and the chimeric quasiperiodic state, respectively.}
    \label{fig5}
\end{figure*}

The $K$-$r$ plots for ring networks of nonlocally coupled oscillators ($p=2$) are shown in Fig.~\ref{fig5}. Here, $p=2$ corresponds to a case where each oscillator is symmetrically coupled to its nearest and next-nearest neighbor on either side of a ring. As compared to the case of local coupling, now the AD region has enlarged, the region of 2-CL and splay state of type 3 has reduced, and more number of splay states (i.e., type 1 and type 2 splay states shown in Figs.~\ref{fig3}a and \ref{fig3}c, respectively) and cluster states (i.e., 3-CL and Gen 4-CL shown in Figs.~\ref{fig2}d and \ref{fig2}e) are observed at low coupling strengths (see zoomed insets for $K<1$ in Fig.~\ref{fig5}). In the case of $N=6$ (Fig.~\ref{fig5}a), a small region of transient chimeric quasiperiodic (CQP) state is also observed at low coupling strengths. However, for the case of $N=8$ (Fig.~\ref{fig5}c), a Gen4CL (Fig.~\ref{fig2}e) state is observed instead of the expected 4-cluster state. In the Gen 4-CL state, a pair of clusters is anti-phase synchronized to each other but has a phase difference of $0^{\circ}$ to $90^{\circ}$ between these pairs (Fig.~\ref{fig2}e). Such a state is observed as the basin of attraction of the Gen-4CL state is larger than the 4-cluster state for general initial conditions \cite{zou_splay_2009}. We also notice that states like AMIP and AM2CL are very short-lived (less than 200 cycles) in the case of nonlocal coupling as compared to local coupling and cannot be distinguished within the range of $K$ and $r$ used in the present study.

\subsubsection{Nonlocal coupling ($p=3$)}
\begin{figure*}
    \centering
    \includegraphics[width = 0.8\textwidth]{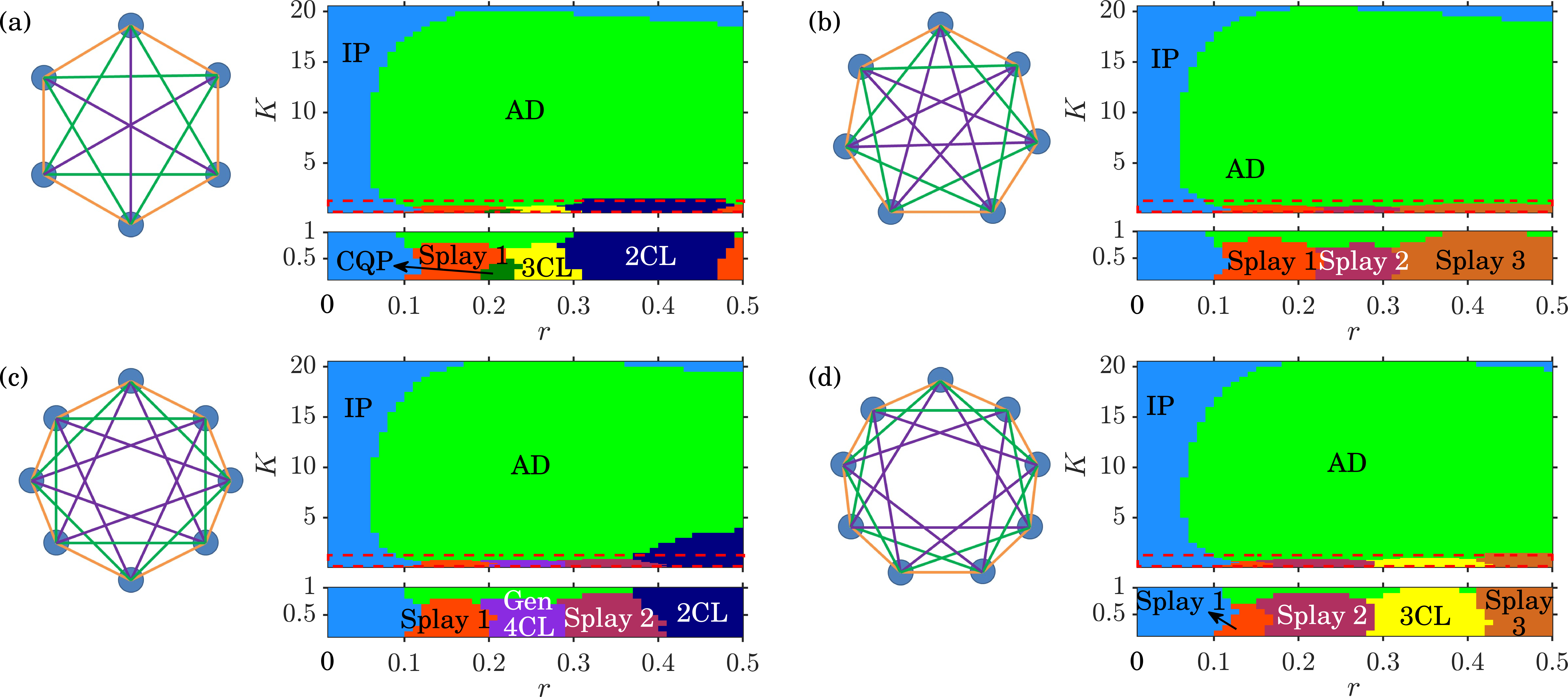}
    \caption{(a)-(d) Pictorial representation of the networks and the corresponding $K$-$r$ plots showing the dynamics of a nonlocally coupled ring of oscillators ($p=3$ case) for $N=6$, $7$, $8$, $9$, respectively. Zoomed inset of the region highlighted by the red dashed box is shown below ($K<1$ region).}
    \label{fig6}
\end{figure*}

In the case of nonlocal coupling with $p=3$, each oscillator is symmetrically connected to 3 neighbors on either side. The $K$-$r$ plots for this case corresponding to different numbers of oscillators ($N$) are shown in Fig.~\ref{fig6}. For all $N$, we observe that the AD region has enlarged and is nearly the same size. The number of splay states observed at low coupling strengths has increased. Transient CQP state is observed at low coupling strengths for $N=6$ (Fig.~\ref{fig6}a) as well. In contrast to the cases of local and nonlocal coupling with $p=2$, we also notice the 2-CL state only at lower coupling strength for an even number of oscillators (Figs.~\ref{fig6}a and~\ref{fig6}c). Moreover, at higher coupling strengths, only IP oscillations are observed, similar to that observed for an odd number of oscillators on the ring (Figs.~\ref{fig4}b, d). The overall dynamical behavior of rings with odd and even numbers of oscillators have become similar, except at low coupling strengths as the occurrence of splay states depends on the number of oscillators in the network.

\subsubsection{Nonlocal coupling ($p=4$)}

\begin{figure*}
    \centering
    \includegraphics[width = 0.8\textwidth]{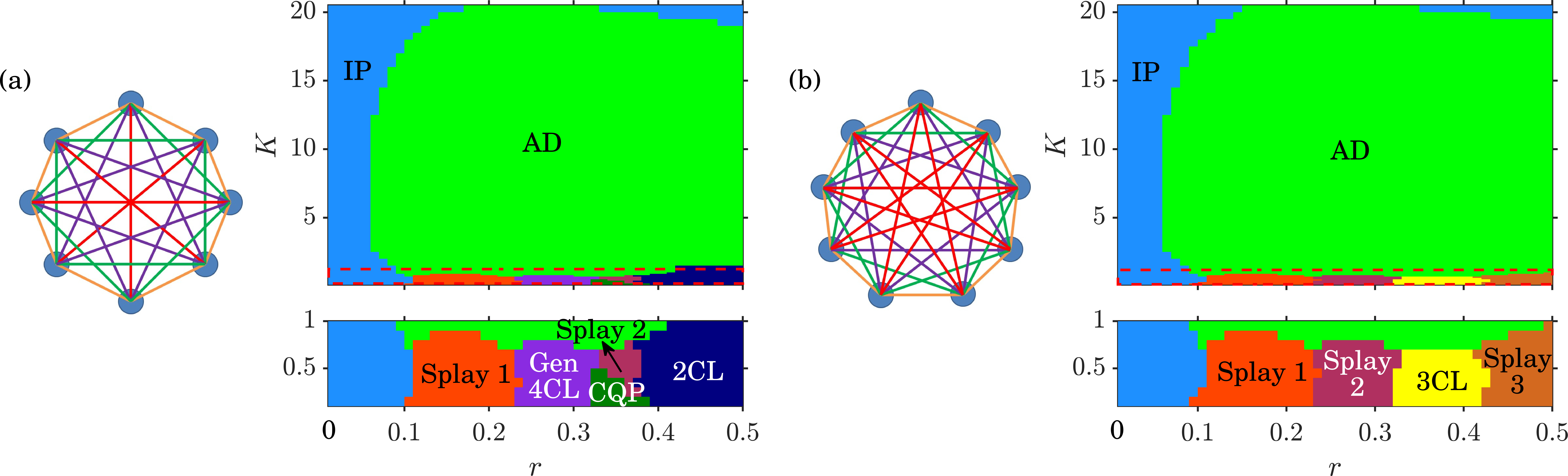}
    \caption{(a)-(d) Pictorial representation of the networks and the corresponding $K$-$r$ plots showing the dynamics of a nonlocally coupled ring of oscillators ($p=4$ case) for $N=6$, $7$, $8$, $9$, respectively. Zoomed inset of the region highlighted by the red dashed box is shown below ($K<1$ region).}
    \label{fig7}
\end{figure*}

In the case of nonlocal coupling with $p=4$, in rings of $N=8$ (Fig.~\ref{fig7}a) and $N=9$ (Fig.~\ref{fig7}b) oscillators, we observe the transient CQP state and all possible splay states at low values of coupling strength ($K<1$). At intermediate coupling strengths, IP oscillations (at lower $r$) and AD states are observed. While at high coupling strengths, only IP synchronization occurs. The overall dynamics are once again nearly similar for both even (Fig.~\ref{fig7}a) and odd (Fig.~\ref{fig7}b) numbers of oscillators.

\subsubsection{Global coupling}
\begin{figure*}
    \centering
    \includegraphics[width = 0.8\textwidth]{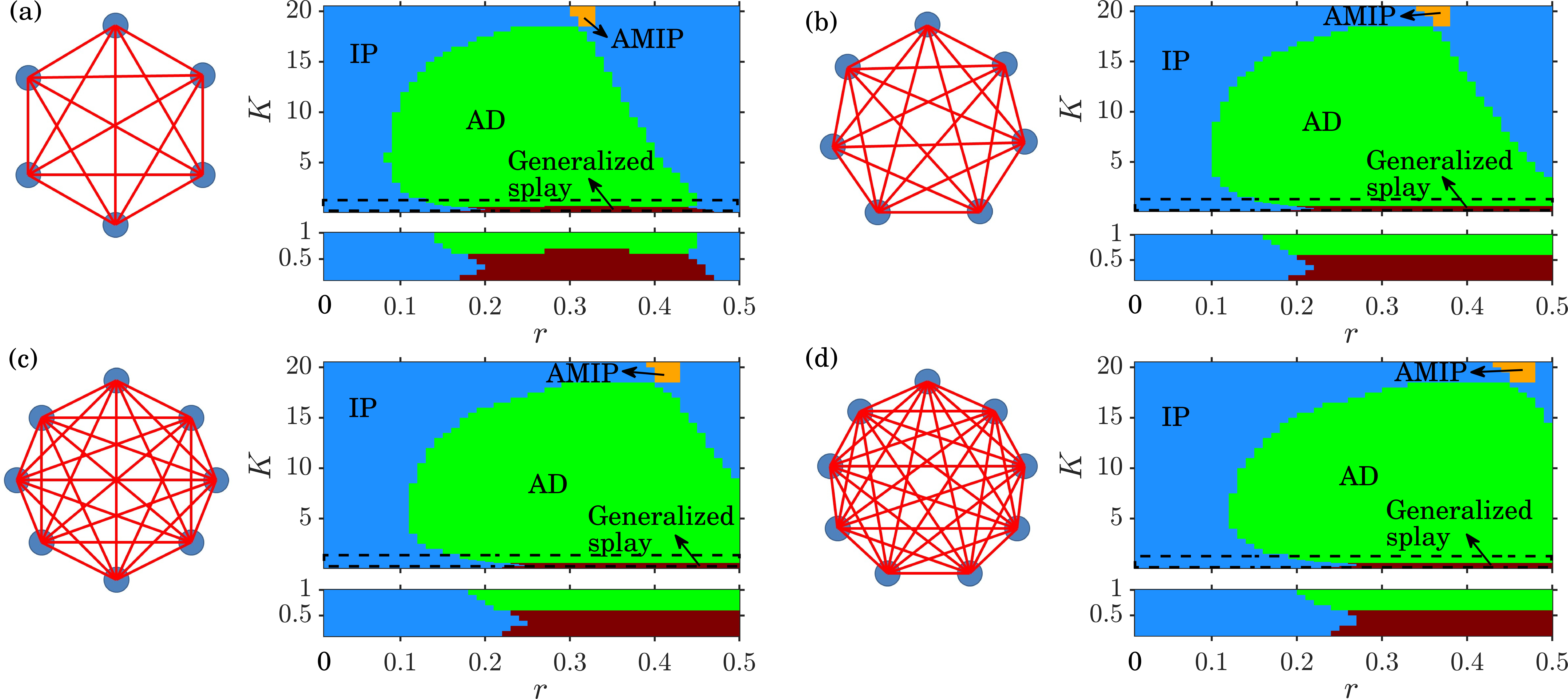}
    \caption{(a-d) Pictorial representation of the networks and the corresponding $K$-$r$ plots showing the dynamics of a globally coupled ring of oscillators for $N=6$, $7$, $8$, $9$, respectively. Zoomed inset of the region highlighted by black dashed box is shown below ($K<1$ region). Here, Generalized splay stands for the generalized splay state.}
    \label{fig8}
\end{figure*}

In the case of globally coupled ring networks, we consider all oscillators to be connected with the same coupling strength and time-delay. We examine this case to better understand how the presence or absence of distant dependent delays influences the dynamics of the ring of oscillators. The $K$-$r$ plots for this case are shown in Fig.~\ref{fig8}. For all $N$, we can see that at low coupling strengths, only IP and generalized splay states are observed. At intermediate coupling strengths, IP and AD occur, while at higher coupling strengths, the IP and AMIP states are observed. In the case of local coupling (Fig.~\ref{fig4}), the AM2CL state is observed when there is an even number of oscillators in the ring networks, while the AMIP state is observed when there is an odd number of oscillators in such networks. In contrast, only the AMIP state is observed regardless of the number of oscillators in the globally coupled oscillators of ring networks. It is also interesting to note that since the generalized splay states have arbitrary phase differences between the oscillators (Fig.~\ref{fig2} f), the number of oscillators does not play a role. Hence, the overall dynamical behavior is independent of the number of oscillators in the case of global coupling. A similar analysis on the globally coupled ring of oscillators was carried out by Reddy \textit{et al.} \cite{ramana_reddy_time_1998}. However, the generalized splay state at low coupling strengths and the transient AMIP behavior at high coupling strengths were not reported in such networks in previous studies. Furthermore, Zou and Zhan \cite{zou_splay_2009} reported the existence of the generalized splay state in a globally coupled ring network of 15 oscillators and delineated the region of stability of this state in the $K$-$a$ space, where $a$ is the Hopf bifurcation parameter. The effect of time-delay on coupled oscillators of a ring was not taken into account in their study. However, our results demonstrate the existence of the generalized splay state in time-delayed globally coupled networks of oscillators for the first time to the best of our knowledge.

\subsubsection{Behavior of amplitude-modulated states}
\sloppy
\begin{figure*}
\centering
\includegraphics[width = 0.8\textwidth]{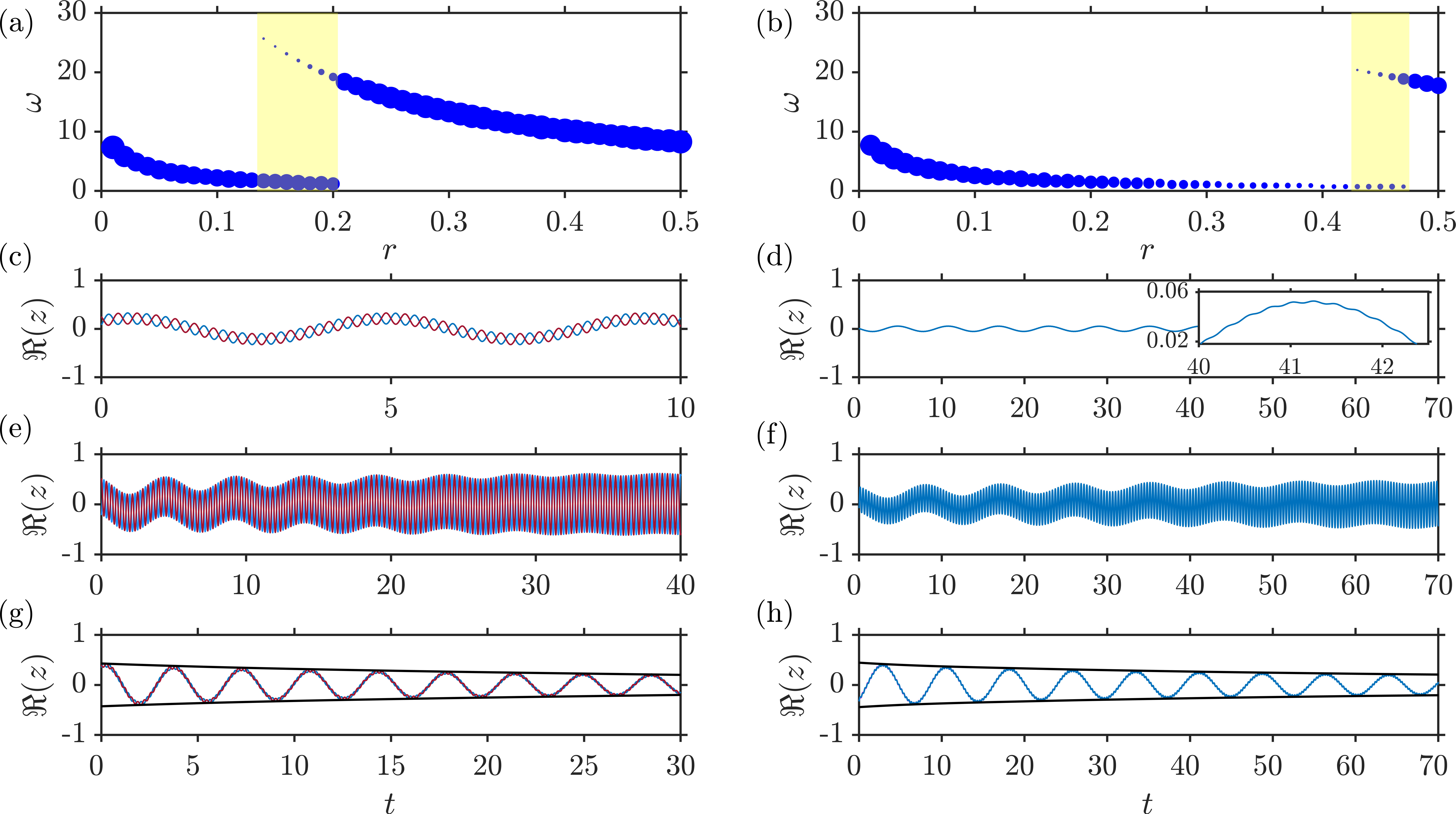}
\caption{Variation of the dominant frequency with $r$ for the (a) amplitude-modulated 2-cluster (AM2CL) state in a ring of $N=8$ oscillators and (b) amplitude-modulated in-phase (AMIP) state in a ring of $N=9$ oscillators, in the case of local coupling, for a fixed $K=20$. The size of each blue dot represents the magnitude of the dominant frequency peaks in the amplitude spectrum of the signal and the yellow regions corresponds to the respective amplitude-modulated state. (c, e, g) Time series of the AM2CL state in a ring of $N=8$ oscillators when (c) it is observed as a persistent state ($K=20$, $r=0.17$), (e) AM2CL state transitions into the 2CL state ($K=20$, $r=0.22$) and (g) the transition from AM2CL to AD state ($K=16$, $r=0.19$) occurs. The amplitude envelope is shown in black. (d, f, h) Time series of the AMIP state in a ring of $N=9$ oscillators, during (d) low amplitude in-phase oscillations, where the larger frequency is smaller in amplitude, such that only upon zooming (see inset), the modulation can be clearly observed ($K=20$, $r=0.32$), (f) AMIP to large amplitude IP transition ($K=20$, $r=0.49$), and (h) AMIP to AD transition ($K=20$, $r=0.41$). The amplitude envelope is shown in black. Initial transients have been removed appropriately.} 
\label{fig9}
\end{figure*} 

\sloppy
We observe two states of amplitude-modulated oscillations, named the amplitude-modulated 2 cluster (AM2CL) state and the amplitude-modulated in phase synchronization (AMIP). They occur in the form of synchronized quasiperiodic oscillations, whose amplitude spectrum is characterized by two incommensurate frequencies that are exactly the same for each oscillator. Figures \ref{fig9}(a) and \ref{fig9}(b) show the variation in the dominant frequencies of oscillators upon varying $r$ while keeping a constant value of $K=20$, in the case of $N=8$ (even oscillators) and $N=9$ (odd oscillators), respectively. We notice a single small frequency component ($\omega<\omega_n$) for low values of $r$ (corresponding to 2CL state in Fig.~\ref{fig9}a and low amplitude IP in Fig.~\ref{fig9}b), and a single large frequency component ($\omega>\omega_n$) for large values of $r$, corresponding to the state of IP oscillations. For some intermediate values of $r$ (yellow shaded regions in Fig.~\ref{fig9}a and \ref{fig9}b) where two frequency components coexist, i.e., one higher and one lower, amplitude-modulated oscillations are observed in the system. 

Figures~\ref{fig9}(d, f, h) show the transition of IP to AMIP, AMIP to IP, and AMIP to AD, respectively, while Figs.~\ref{fig9}(e, g) show the transition of the AM2CL state to the 2CL state and AD state, respectively. Interestingly, in the case of the AM2CL state observed for values of $r$ in the middle of the AM2CL region, the state is highly stable, as shown in Fig.~\ref{fig9}(c). However, the AMIP state is, in comparison to AM2CL, short-lived for all values of $r$. It is interesting to note from Fig.~\ref{fig9}(b) that there is a jump in the amplitude of the IP oscillations across the AMIP state as $r$ is varied. This is evident from the size of the dots, representing the height of the frequency peak in the amplitude spectrum (Fig.~\ref{fig9}a, b). This observation is different from the AM2CL state, where the amplitudes of the IP and 2CL states before and after the AM2CL region respectively, are comparable to each other.

Hence, from the above discussion, it is clear that the occurrence and lifetime of transient states, such as the AMIP and AM2CL, critically depend on the coupling parameters such as the time-delay, coupling strength, coupling scheme, and the number of oscillators in ring networks. Haugland \textit{et al.} \cite{haugland_between_2021,haugland_coexistence_2023} showed the existence of a stable state resembling the AM2CL oscillations, albeit in a model involving nonlinear global coupling. To the best of the authors' knowledge, the present work is the first to explore the dependence of the lifetime of such a state on the coupling scheme, coupling strength, and time-delay in the network. 

\subsection{\label{chain_star}Dynamical behavior of chain and star networks}

In this section, we present the coupled behavior of oscillators in other minimal networks with open topologies, such as chain and star networks, and compare their dynamics with those observed in ring networks discussed previously. 

\subsubsection{Chain topology}
\begin{figure*}
    \centering
    \includegraphics[width = 0.85\textwidth]{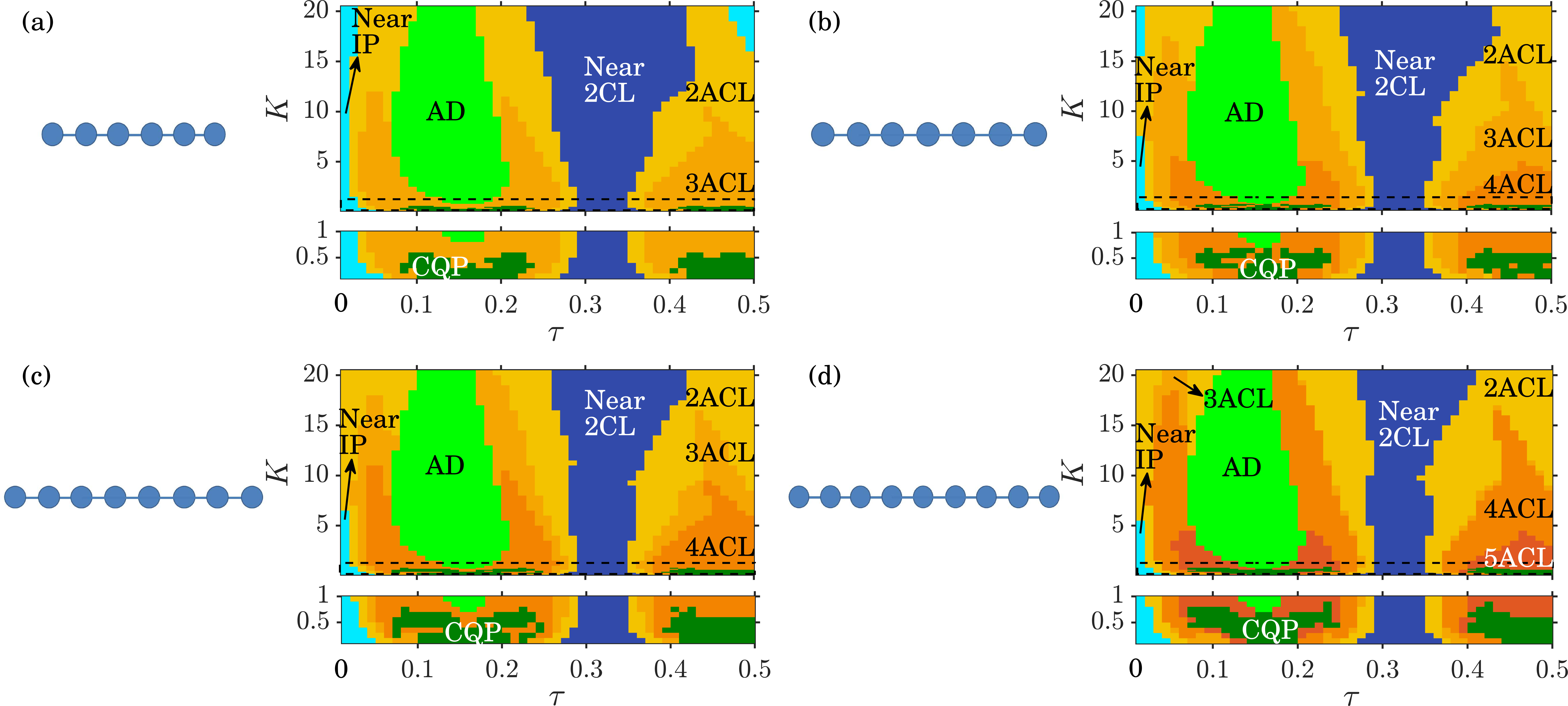}
    \caption{(a)-(d) Pictorial representation of the network and $K$-$\tau$ plot showing the dynamics of oscillators coupled in a chain for $N=6$, $7$, $8$, $9$, respectively. Zoomed inset of the region highlighted by black dashed box is shown below ($K<1$ region). Here, Near IP, Near 2CL and ACL denote the near in-phase state, near 2-cluster state and amplitude clustering state, respectively.}
    \label{fig10}
\end{figure*}

The $K$-$\tau$ plots for oscillators coupled in a chain topology are shown in Fig.~\ref{fig10}. We observe near IP state for low values of $r$ (Fig.~\ref{fig2}g) and a near 2-CL state (Fig.~\ref{fig2}g) for intermediate values of $r$. The near 2-CL states are also observed for an odd number of oscillators, unlike in a ring topology, where similar 2-cluster states are observed only with an even number of oscillators. We observe an amplitude death region at intermediate time-delays, whose size remains nearly the same for chain networks with different numbers of oscillators, contrary to that observed in the ring topology, in which case the size of the death island remarkably depended on the number of oscillators in the ring. We also observe regions of amplitude clusters (ACL), where the oscillators form groups having different amplitudes but are synchronized (Fig.~\ref{fig1}e). The number of amplitude clusters formed increases as the coupling strength $K$ decreases. To the best of our knowledge, this is the first observation of amplitude clustering in a time-delay coupled system along with a detailed description of the trend in its occurrence as a function of coupling parameters. Further, we observe the transient CQP state at low coupling strengths for all values of $N$.

\subsubsection{Star topology}

\begin{figure*}
    \centering
    \includegraphics[width = 0.85\textwidth]{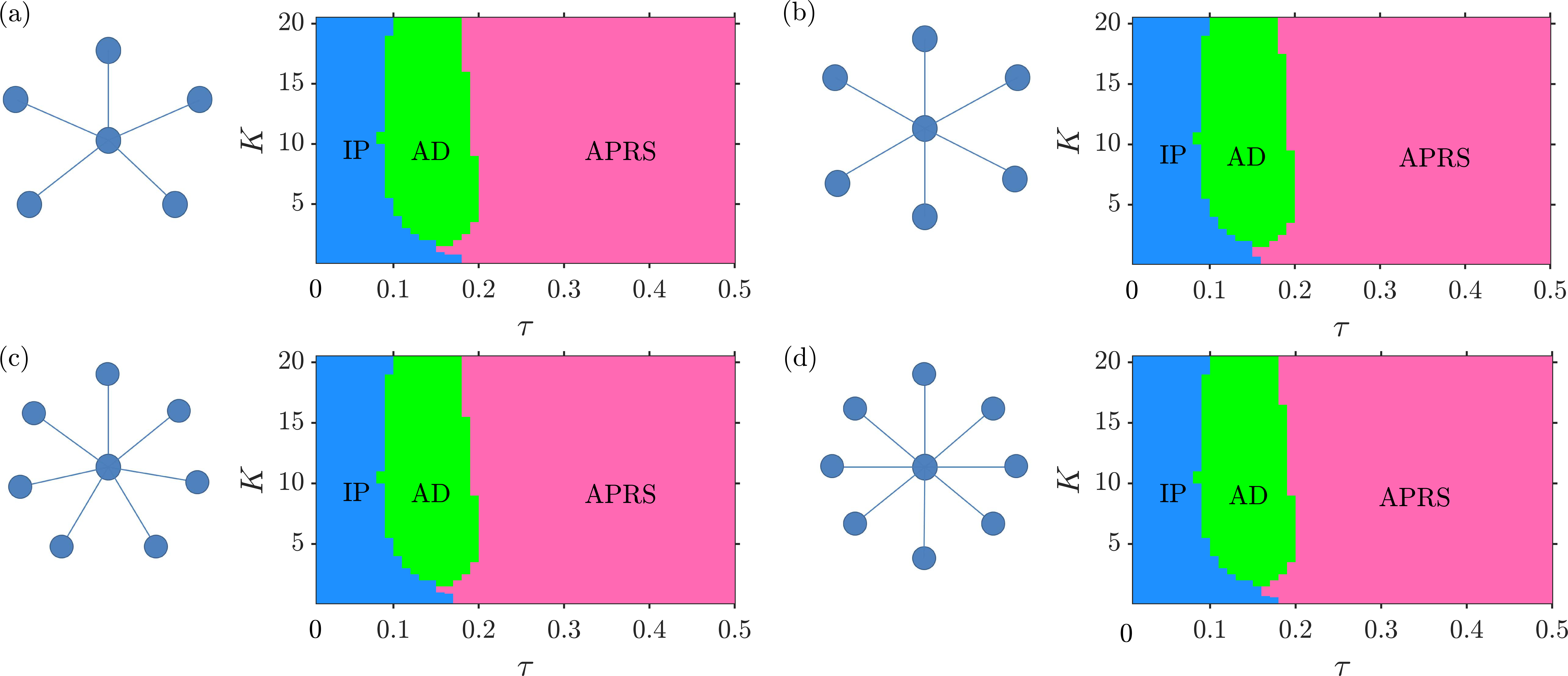}
    \caption{(a)-(d) Pictorial representation of the networks and the corresponding $K$-$\tau$ plots showing the dynamics of oscillators coupled in a star network for $N=6$, $7$, $8$, $9$, respectively. Here, IP, AD and APRS stand for in-phase state, amplitude death, and anti-phase remote synchronization, respectively.}
    \label{fig11}
\end{figure*}

The $K$-$\tau$ plots for oscillators in a star network topology are shown in Fig.~\ref{fig11}. Here, we observe that for any number of oscillators, in-phase synchronization occurs for lower time-delay, amplitude death for intermediate time-delay, and anti-phase remote synchronization (APRS) \cite{chen_stability_2022} for higher time-delays. As discussed in Fig.~\ref{fig1}(e,f), the APRS state is characterized by the hub node being anti-phase synchronized to the leaf nodes, which are in-phase synchronized to each other remotely via the hub node. Such states could be observed in the study of brain networks \cite{nicosia_remote_2013} and climactic interactions between different regions \cite{bergner_remote_2012} where long-range connections play an important role. The amplitude death region is observed to have the same size for all $N$. As in the case of globally coupled ring networks or chain networks, the dynamics of the star network mostly does not depend on the number of oscillators in the network.

\section{\label{conclusion}Conclusions}
\sloppy
In this paper, we systematically investigated the dynamics of minimal networks of time-delay coupled Stuart-Landau oscillators under the variation of different network parameters such as the number of oscillators, coupling schemes, and network topologies. We varied the number of oscillators from 6 to 9 to investigate the effect of odd and even numbers of oscillators in the network. Furthermore, we examined three network topologies including ring, star, and chain, where the first topology is a close-loop structure and the last two topologies are open-loop structures. In a ring, we also studied the effect of change in the coupling of oscillators from local to non-local to global. 

We observed that in the case of ring networks with locally coupled oscillators, the overall dynamical behavior has stark differences depending on whether there are an odd or even number of oscillators in the network. For both cases, we noticed four major dynamical phenomena as in-phase synchronization, clusters, splay states, and amplitude death. Additionally, novel amplitude-modulated transient behaviors such as the amplitude-modulated in-phase (AMIP) and the amplitude-modulated 2-cluster (AM2CL) have also been observed at high coupling strength and intermediate time-delay. The AM2CL state is, in general, a transient, but can last for more than 5000 cycles depending on the coupling strength and time-delay. It is characterized by in-phase synchrony at long time scales and anti-phase synchrony at short time scales. On the other hand, the AMIP state is much shorter lived, and occurs in the transition from low amplitude in-phase to high amplitude in-phase oscillations as the time-delay is increased. Such amplitude-modulated states could be of practical relevance in studying ecological dynamics \cite{defriez_climate_2016} and thermoacoustic oscillations in combustors \cite{roy_flame_2021}. The occurrence and lifetime of these transient states are critically dependent on the network properties, such as the coupling scheme, coupling strength, and time-delay. 

As the nonlocality in the coupling structure of a ring network is increased, complex dynamics are observed at lower coupling strengths. The number of different splay states and cluster states observed increases at low coupling strengths. We also showed the existence of transient weak chimera in nonlocally coupled ring networks and the generalized splay state in globally coupled ring networks. Moreover, we noticed that as the number of connections to an oscillator increases and as the coupling structure transitions to the globally coupled ring of oscillators, the overall dynamical behavior of the network appears to become independent of the number of oscillators.

As the network topologies changed from ring to open network structures such as the chain and the star, the dynamical behavior of coupled oscillators displayed a significant change at lower coupling strength and higher coupling delays. Interestingly, we notice that chain networks induce both phase shifts and amplitude variations among the synchronized oscillations of the oscillators. Thus, it appears that chain networks are reluctant to undergo perfect synchronization as seen in the dynamics of ring networks. In this network, we observed amplitude clustering behavior as well as desynchronized quasiperiodic oscillations. Furthermore, star networks demonstrated in-phase synchronization, amplitude death, and a unique state of anti-phase remote synchronization for any number of oscillators. 

Therefore, the results obtained from the present study can provide deeper insights into the complex behavior of oscillators in minimal networks encountered in different fields of science and engineering, ranging from aeroelastic flutter \cite{raaj_investigating_2021} to dynamics of hair bundles in the inner ear \cite{kim_amplitude_2014}. Yet another possible application of our work is understanding the dynamics of coupled can-annular and annular combustors used in gas turbines \cite{fang_experimental_2021,farisco_thermo-acoustic_2017,moon_modal_2023,pedergnana_coupling-induced_2022}, which are utilized in power generation and aerospace propulsion. These combustors are prone to thermoacoustic instability, characterized by large-amplitude, self-sustained limit cycle oscillations in the system \cite{sujith_thermoacoustic_2021}. These instabilities primarily occur due to positive feedback between the unsteady heat release rate oscillations in the flame and the acoustic field in a combustor \cite{lieuwen_combustion_2006}. The dynamics in each burner (or can) of the combustor during thermoacoustic instability can be modeled as a limit cycle oscillator. In a can-annular combustor, such oscillators are coupled to each other by the acoustic field transmitted through the inter-connectors, also called cross-fire tubes \cite{farisco_thermo-acoustic_2017}, or at the turbine inlet \cite{pedergnana_coupling-induced_2022}. In the case of an annular combustor, the flames primarily interact with their neighbors through hydrodynamic and acoustic fluctuations in the reaction flow field \cite{fang_experimental_2021}.

Depending on which mode of thermoacoustic instability is excited in the system, the acoustic field may be coupled to the flames differently. For example, in the case of longitudinal instability, the acoustic field is globally coupled to all flames in a similar manner, and only in-phase synchronized oscillations in flames occur \cite{roy_flame_2021}. However, in azimuthal thermoacoustic instability, the acoustic field rotates in the azimuthal direction, and the coupling of individual flames with the acoustic field varies with time \cite{evesque_spinning_2003}. This results in several other synchronization phenomena, such as standing and spinning waves \cite{faure-beaulieu_imperfect_2021}. The finite propagation timescales of acoustic pressure and heat release rate fluctuations induce delays in the interaction between the oscillators \cite{thomas_effect_2018,moon_mutual_2020}. Studying the actual dynamics in such combustors requires large-scale test rigs with multiple burners, which are costly and challenging to operate. Another alternative is to perform high-fidelity numerical simulations using coupled governing equations of flow-flame acoustic fluctuations \cite{fournier_low-order_2021,moon_modal_2023}; however, these are time-consuming and costly. Hence, we believe that reduced-order models based on coupled Stuart-Landau oscillators \cite{premraj_effect_2021}, as discussed in the present study, are indispensable tools for possibly anticipating and controlling the dynamics of practical can or annular type combustors. 

\begin{acknowledgements}
A.E.B. thanks Prof. Oded Gottlieb (Technion -- Israel Institute of Technology) for useful discussions during the preparation of this manuscript.
\end{acknowledgements}

\section*{Statements and Declarations}

\subsection*{\textbf{Funding}} 
The authors acknowledge the financial support from the J. C. Bose Fellowship (ASE/18-19/169/SERB/RISU) from the Science and Engineering Research (SERB) of the Department of Science and Technology (DST) of the Government of India, and from the Institute of Eminence (IOE) initiative (No. SB/2021/0845/AE/MHRD/002696) from the Ministry of Human Resources and Development (MHRD) of the Government of India.

\subsection*{\textbf{Conflict of interest/Competing interests}} 
The authors have no relevant financial or non-financial interests to disclose.

\subsection*{\textbf{Availability of data and materials}} 
The datasets generated and/or analyzed during the current study are available from the corresponding author on reasonable request.

\subsection*{\textbf{Ethics approval}} 
Not applicable.

\subsection*{\textbf{Consent to participate}} 
Not applicable.

\subsection*{\textbf{Consent for publication}} 
Not applicable.

\subsection*{\textbf{Code Availability}} 
The codes used in the current study are available from the corresponding author on reasonable request.

\subsection*{\textbf{Authors' contributions}} 
All authors contributed to the study conception and design. Material preparation, data collection and analysis were performed by Andrea Elizabeth Biju and Sneha Srikanth. The first draft of the manuscript was written by Andrea Elizabeth Biju and all authors commented on previous versions of the manuscript. All authors read and approved the final manuscript.

\bibliographystyle{spphys}       
\bibliography{references}   

\end{document}